\documentclass[seceq]{ptptex}
\usepackage{graphicx}
\usepackage{wrapft}

\title{Systematic Survey of the Correlation between Northern HECR Events and SDSS Galaxies}

\author{
Hajime \textsc{Takami}$^{1}$, 
Takahiro \textsc{Nishimichi}$^{2}$, 
and Katsuhiko \textsc{Sato}$^{2,3}$}

\inst{
$^{1}$Max Planck Institute for Physics, F\"{o}hringer Ring 6, 80805 Munich, Germany \\
$^{2}$Institute for the Physics and Mathematics of the Universe, the University of Tokyo, 5-1-5, Kashiwanoha, Kashiwa, Chiba 277-8583, Japan \\
$^{3}$National Institutes of Natural Sciences, 4-3-13, Toranomon, Minato-ku, 105-0001 Tokyo, Japan
}

\abst{We investigated the spatial correlation between the arrival directions of the highest energy cosmic rays (HECRs) detected by the Akeno Giant Air Shower Array (AGASA) with energies above $4 \times 10^{19}$ eV and the positions of galaxies observed by the Sloan Digital Sky Survey (SDSS) within $z = 0.024$. We systematically tested the dependence of the correlation on the redshift ranges and properties of the galaxies, i.e., absolute luminosity, color, and morphology, to understand where HECR sources are and what objects are HECR sources. In the systematic survey, we found potential signals of the positive correlation at small angular scale ($<10^{\circ}$) with the (non-penalized) chance probability less than 5\% in intermediate redshift ranges. Then, we estimated penalized probabilities to compensate the trial effects of angular scan, and found that the strongest correlation is produced by early-type galaxies in $0.012 \leq z < 0.018$ at 90\% C.L. The possible origin of HECRs which correlating galaxies imply is also discussed.}

\PTPindex{411}

\begin{document}

\maketitle

\section{Introduction} \label{introduction}

The origin of the highest energy cosmic rays (HECRs) is an intriguing problem in astrophysics. It is widely believed that HECRs are of extragalactic origin because they cannot be confined in the Galaxy by a Galactic magnetic field (GMF) and their arrival direction distribution does not correlate with Galactic plane. Plausible source candidates of HECRs ever discussed are $\gamma$-ray bursts (GRBs),\cite{waxman95,vietri95,murase06} \ radio-loud active galactic nuclei (AGN),\cite{takahara90,rachen93,berezhko08,farrar09,takami11} \ radio-quiet AGN,\cite{pe'er09} \ magnetars,\cite{arons03,murase09} \ and clusters of galaxies.\cite{kang97,inoue06} \ The sources of cosmic rays with $\sim 10^{20}$ eV detected at the Earth are limited to objects with distances less than 100 Mpc because of interactions of HECRs with cosmic microwave background (CMB) photons, so-called Greisen-Zatsepin-Kuz'min (GZK) mechanism.\cite{greisen66,zatsepin66} \ Thus, it is expected that the arrival directions of HECRs correlate with the directions of their sources if the deflections of HECR trajectories by a GMF and intergalactic magnetic fields (IGMFs) are not large. The Akeno Giant Air Shower Array (AGASA) reported anisotropy in the arrival distribution of HECRs at small angular scale, which is comparable with the accuracy of the determination of the arrival directions of HECRs.\cite{takeda99} \ This anisotropy has been thought to be appearance of nearby astrophysical sources, but HECR sources have not been clear yet.

Important progress was brought in by the Pierre Auger Observatory (PAO). The PAO reported a correlation between the arrival directions of 27 cosmic rays with energies above $\sim 5.7 \times 10^{19}$ eV and the positions of galaxies listed in the 12th Veron-Cetty and Veron (VC) catalog\cite{veron06} with $z \leq 0.018$ within the angular scale of $\sim 3.1^{\circ}$.\cite{abraham07,abraham08} \ Although the VC catalog is for AGN, the original argument of the PAO conservatively claimed that the distribution of cosmic ray sources is correlated with matter distribution in local Universe involving AGN because the spatial distribution of the underlying matter density correlates with that of AGN, i.e., the AGN listed in the VC catalog are just tracers of the matter distribution.\cite{abraham07} \ Several groups have also analyzed the PAO data by using catalogs of different astrophysical objects: X-ray selected AGN,\cite{george08} \ HI-selected galaxies,\cite{ghisellini08} \ and infrared-selected galaxies,\cite{kashti08,takami09c} \ and they all reported significant correlations. Thus, the PAO data indicates that HECR sources are distributed to being comparable with the matter distribution in local Universe. Although the number of detected events is increasing after the first publication, the significance is fluctuating between $2\sigma$ and $3\sigma$.\cite{hague09,aublin09}

The PAO can observe only the southern sky because it is located in Argentina. Is the correlation feature universal even in the northern sky? There are two reasons why the answer of this question is not trivial. The first reason is the structure of a GMF. Although its impact depends on GMF models, several models indicate that the typical deflection angles of HECR trajectories in Galactic space is different between in the northern sky and in the southern sky.\cite{takami08b,takami09f} \ The second reason is the results of composition measurements. The measurements of $\left< X_{\rm max} \right>$ and its fluctuation by the PAO indicate heavy nuclei dominated composition at the highest energies,\cite{abraham10} \ while those by the High Resolution Fly's Eye (HiRes), which is a fluorescence detector in the northern hemisphere, imply proton-dominated composition.\cite{abbasi05,abbasi10b} \ Although this is a controversial issue, if this difference is intrinsic, HECRs detected in the northern sky experience smaller deflections and could produce stronger correlation.

The correlation between HECRs and the matter distribution in the northern sky has been discussed. Refs.~\citen{stanev95} and \citen{uchihori00} reported the correlation of HECRs with the supergalactic plane. We also examined the correlation of the AGASA data with galaxies listed in Infrared Astronomical Satellite Point Source Redshift Survey (IRAS PSCz) catalog,\cite{saunders00} \ but did not find significantly positive correlation.\cite{takami09c} \ Recently, the HiRes reported no significant correlation based on the same analysis as the PAO using its own data.\cite{abbasi08b} \ The HiRes also tested correlation between its data and galaxies observed by 2 Micron All-Sky Redshift Survey (2MASS),\cite{abbasi10} \ but they reported no significant correlation at 95\% C.L..

In this study, we test the correlation between the AGASA events as a representative of HECR event sets detected in the northern sky and galaxies observed by the Sloan Digital Sky Survey (SDSS).\cite{york00} \ The SDSS observes even fainter galaxies down to $m_r \sim 20.0$, where $m_r$ is the $r$-band apparent magnitude of galaxies, and gives us dense galaxy samples, which can well reflect galaxy distribution at small angular scale even by using volume-limited samples. A SDSS sample of galaxies used in this study has more than 10,000 galaxies within $\sim 100$ Mpc ($z = 0.024$) after cuts for a volume-limited sample (see Section \ref{galaxies}) in spite of its small sky coverage ($\sim 20$\% of the entire sky). For comparison, note that the 2MASS sample used by the HiRes has 15,508 galaxies in the HiRes field of view (about a half of the whole sky) within 250 Mpc. In addition, the completeness of the observed sky is well studied, which enables us to estimate the geometry of the galaxy survey precisely.

Recent studies on the correlation between HECRs and galaxies in astronomical catalogue have regarded the galaxies as representatives of the matter distribution in local Universe. However, the statistical properties of the distribution of galaxies are, in general, different among the types of galaxies, which has been well studied in the field of cosmology. This can be easily understood when rare objects such as strong radio galaxies are considered. This viewpoint is useful to understand HECR sources and many works have been dedicated to test the correlation between HECRs and source candidates, i.e., different types of galaxies. Refs.~\citen{smialkowski02} and \citen{singh04} studied the correlation with infrared galaxies using the IRAS PSCz catalog of galaxies\cite{saunders00} and discussed the possibility that the AGASA events below $\sim 10^{20}$ eV arrive from luminous infrared galaxies. The correlation of the AGASA and Yakutsk data with BL Lac objects was also suggested.\cite{tinyakov01,tinyakov02,gorbunov02} \ AGN in the Rossi X-ray Timing Explorer (RXTE) catalog was tested, but do not correlate with HECRs significantly.\cite{hague07} \ Ref.~\citen{gorbunov05} performed a comprehensive study of the correlation with various classes of extragalactic objects and pointed out that BL Lac objects and unidentified gamma-ray sources might correlate with the AGASA and Yakutsk data.

Following this standpoint, we focus on the correlation with galaxies, not the underlying matter distribution. We examine the dependence of the correlation on the properties of galaxies systematically to investigate what objects correlate with HECRs. Since galaxies hosted by HECR sources have characteristic features, which depends on source population (discussed in Section \ref{discussion}), the dependence on the properties of galaxies correlating with HECRs gives useful information on HECR sources. Quantities attached to a SDSS catalog of galaxies allow to calculate useful astronomical quantities of galaxies, e.g., color and morphology of galaxies.

A volume-limited sample of galaxies lacks some information on darker galaxies than a cut criterion. Ref.~\citen{koers09} proposed a useful method to construct a flux reference map of HECRs with compensating this effect. In this method, weights are given to visible galaxies instead of galaxies invisible due to their darkness to correct selection effects by a flux limit. However, the weights does not consider the properties of invisible galaxies. Since we want to examine the dependence of the correlation on the properties of galaxies in this study, we analyze the HECR data by using volume-limited samples instead of this method. Thus, since we need dense galaxy samples to resolve galaxy distribution even at small angular scale, a SDSS catalog of galaxies is appropriate for this study, despite that about a half of the AGASA events does not overlap the field surveyed by the SDSS.

This paper is laid out as follows: the details of HECR events and galaxies used for analysis, and a statistical method to test the correlation between them are described in Section \ref{methods}. In Section \ref{results}, the results of our statistical analysis are presented. Firstly, we test the correlation of HECR events and galaxies in four redshift ranges to investigate where HECR sources are in Section \ref{redshift}. Next, we examine the dependence of the correlation on the properties of galaxies and search for the potential signals of positive correlation in the following two sections. Then, we compensate statistical penalties and estimate the true significance of the positive correlation in Section \ref{penalty}. Section \ref{discussion} is devoted to discuss the relation between HECR source candidates and the properties of their host galaxies.  Finally, we conclude this study in Section \ref{conclusion}.

\section{Data Samples and Statistical Method} \label{methods}

\subsection{Sample of highest energy cosmic rays} \label{events}

As a sample of HECR events detected in the northern sky, the AGASA events listed in Ref.~\citen{hayashida00} are adopted. The AGASA event is the largest HECR sample detected in the northern sky in which the energy and arrival direction of each event are published. Although the total exposure of the HiRes is larger than that of the AGASA, the HiRes has not published the properties of each event. In addition, the AGASA is a ground array, and therefore its duty cycle is approximately 100\%. This fact allows us to estimate the directional dependence of its aperture analytically.

The AGASA sample consists of 57 events with energies above $4 \times 10^{19}$ eV. These events are distributed in the sky of $-10^{\circ} \leq \delta \leq 80^{\circ}$, where $\delta$ is the declination of the arrival direction of each cosmic ray. 8 events out of the 57 events have energies above $10^{20}$ eV, which lead to the extension of cosmic ray spectrum beyond the predicted GZK cutoff.\cite{takeda98} \ The excess events over the GZK cutoff have sometimes been thought to originate from new physics beyond the standard model (see Ref. \citen{bhattacharjee00} for a review), but the HiRes and PAO have recently reported HECR spectra with a GZK-like feature,\cite{abbasi08,abraham08b} \ which indicate astrophysical origin of HECRs. Although the reason of the excess events is still unclear, we assume that all the AGASA events are of astrophysical origin and use them in this paper. 

\subsection{Sample of galaxies} \label{galaxies}

\begin{wrapfigure}{l}{6.6cm}
\centerline{\includegraphics[width=0.98\linewidth]{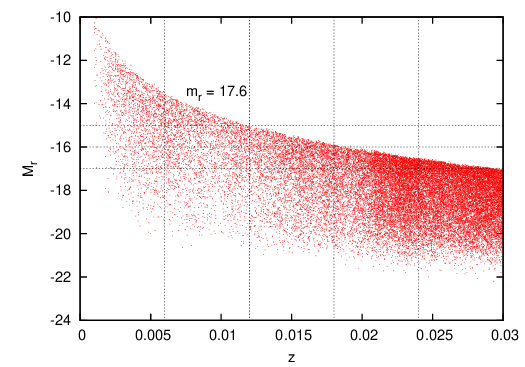}}
\caption{Distribution of SDSS {\it dr72bright0} sample of galaxies with redshifts less than 0.03 on $z$-$M_r$ plane. The upper bound of the distribution corresponds to $m_r = 17.6$. Additional dotted lines are $z = 0.006$, $0.012$, $0.018$, $0.024$, and $M_r = -15$, $-16$, $-17$ for practical use in Section \ref{galaxies}. $M_r$ and $m_r$ are the $r$-band absolute magnitude of galaxies and the $r$-band apparent magnitude of galaxies, respectively.}
\label{fig:z2Mr}
\end{wrapfigure}

Volume-limited samples of galaxies used in this study are constructed from the 7th data release (DR7) of the SDSS,\cite{abazajian09} \ especially large-scale structure subsample named {\it dr72bright0} sample of the New York University Value Added Catalog (NYU-VAGC).\cite{blanton05} \ This is a spectroscopic sample of galaxies with $u$,$g$,$r$,$i$,$z$-band (K-corrected) absolute magnitudes, $r$-band apparent magnitude $m_r$, redshift, an indicator of the morphology, and information on the mask of the survey. These spectroscopic galaxies are selected by the conditions of $10.0 \leq m_r \leq 17.6$ and $0.001 \leq z \leq 0.5$. In fact, the upper bound of apparent magnitude of galaxies is less than 17.6 in a very small fraction of regions, but it is practically no problem to regard the upper bound as 17.6 because the fraction of such regions is less than 0.1\%. The flux limit $m_r = 17.6$ restricts galaxies observed at each redshift, i.e., the maximum $r$-band absolute magnitude $M_r$ of observed galaxies depends on their redshifts. In order to avoid this dependence, we construct the samples of galaxies limited by a constant $M_r$ within redshift which we are interested in, so-called volume-limited samples. We focus on galaxies within $\sim 100$ Mpc in this study, which corresponds to $z \sim 0.024$ in the standard $\Lambda$CDM cosmology. Fig.~\ref{fig:z2Mr} plots all the galaxies within $z \leq 0.03$ in $z$-$M_r$ plane. We can clearly observe a curve due to the flux limit determined by $m_r = 17.6$. This figure indicates that a set of galaxies with $M_r \leq -17$ can be a volume-limited sample within $z = 0.024$, Thus, galaxies with $M_r \leq -17$ are mainly considered in this paper.

In order to constrain the sources of HECRs, we construct subsamples of the volume-limited sample and examine the dependence of the spatial correlation between HECRs and galaxies on the redshift, absolute magnitude, color, and morphology of galaxies. At first, the original volume-limited sample is simply divided into four subsets in redshift for every 0.006, which corresponds to every $\sim 25$ Mpc, to find where HECR sources are. It will be shown in Section \ref{discussion} that HECR flux to which sources in each shell contribute are comparable if the sources are distributed uniformly. Analyzing these four subsets, we will find potential positive correlation signals for galaxies within $0.006 < z \leq 0.012$ and $0.012 < z \leq 0.018$. Following this result, we focus on galaxies in these two redshift ranges and test the dependence on the other parameters as next steps.

Then, two subsets of the four in $0.006 < z \leq 0.012$ and $0.012 < z \leq 0.018$ are divided into two subsets with $-19 < M_r \leq -17$ and $M_r \leq -19$ to test the dependence of the correlation on $M_r$. In addition, galaxies with $-17 < M_r \leq -15$ for $0.006 < z \leq 0.012$ and galaxies with $-17 < M_r < -16$ for $0.012 < z \leq 0.018$ are also adopted as other volume-limited samples to take fainter galaxies into account (see Fig.\ref{fig:z2Mr}).

\begin{figure}
\begin{center}
\includegraphics[width=0.48\linewidth]{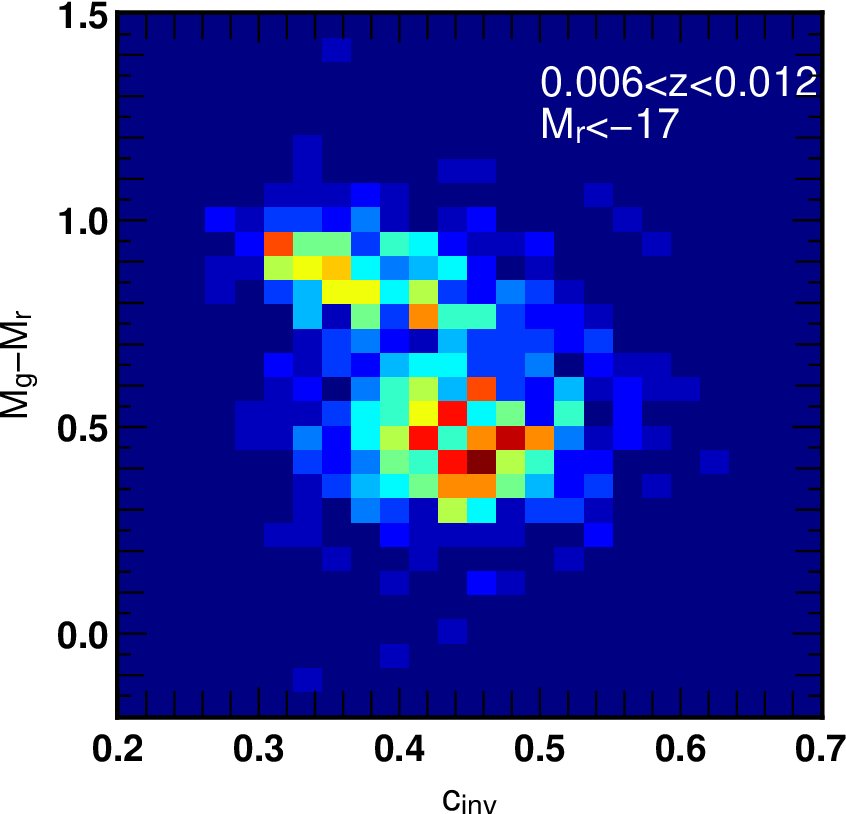} \hfill
\includegraphics[width=0.48\linewidth]{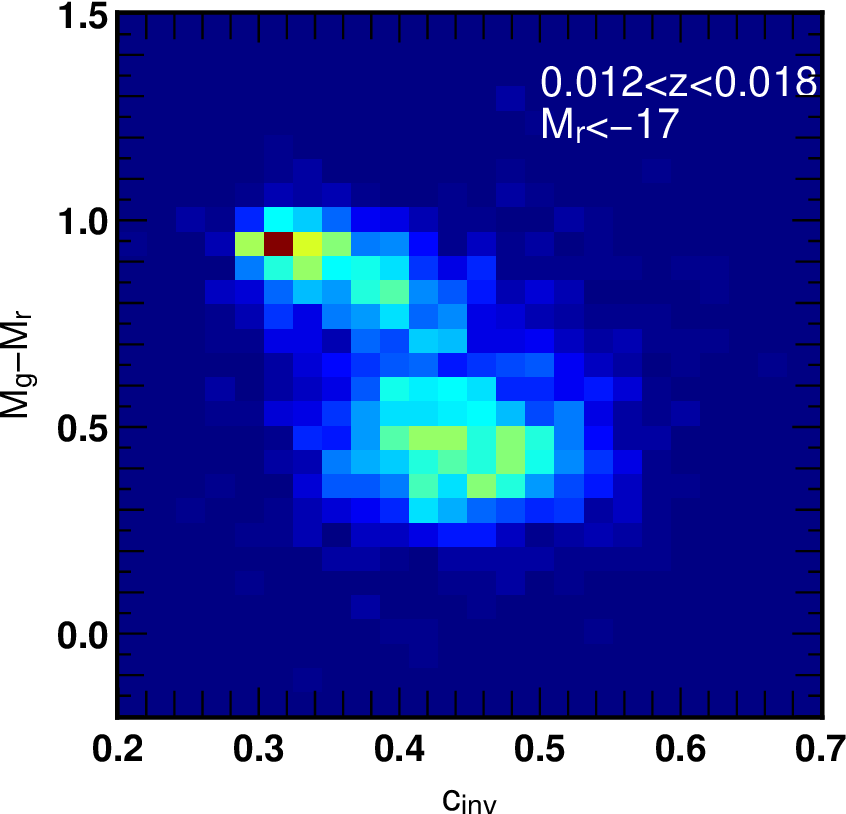}
\caption{Number distribution of SDSS galaxies with the redshifts of $0.006 \leq z < 0.012$ ({\it left}) and $0.012 \leq z < 0.018$ ({\it right}) on $c_{\rm inv}$-$M_g - M_r$ plane. $c_{\rm inv}$ and $M_g - M_r$ are the concentration parameter
of galaxies (see text) and a color indicator used in this study, respectively. The number of galaxies increases from blue to red.}
\label{fig:grcinv}
\end{center}
\end{figure}

The color of a galaxy is generally defined as the difference between magnitudes in 2 different bands in astronomy. Here, the colors of galaxies are defined as $M_g - M_r$, where $M_g$ is $g$-band absolute magnitude of galaxies. In order to estimate the morphology of galaxies, we adopt the ratio of two Petrosian radii $c_{\rm inv}=r_{50}/r_{90}$ called a (inverse) concentration index listed in the NYU-VAGC as an indicator of the morphology. Since early-type (elliptical) galaxies tend to fade out slowly with distance from the center, they have relatively smaller concentration indices. On the other hand, late-type (i.e., spiral and irregular) galaxies have large concentration indices. Fig. \ref{fig:grcinv} shows the frequency distributions of galaxies with $M_r \leq -17$ and the redshift ranges of $0.006 \leq z < 0.012$ ({\it left}) and $0.012 \leq z < 0.018$ ({\it right}) on parameter space of $c_{\rm inv}$ and $M_g - M_r$. The number of galaxies increases when the color in each cell is close to red. In both panels, the distribution of galaxies can be separated above and below $M_g - M_r \sim 0.6$. Thus, we define galaxies with $M_g - M_r$ above and below 0.6 as red and blue galaxies, respectively. On the other hand, these distributions cannot be clearly separated in the direction parallel to $c_{\rm inv}$. Ref.~\citen{shimasaku01} compared the concentration indice SDSS galaxies with apparent magnitude below $16.0$ detected during a commissioning phase to their morphologies and found that the best criterion between early-type and late-type morphologies is $c_{\rm inv} \sim 0.35$. We simply adopt this criterion for classification between early-type galaxies and late-type galaxies.

All the volume-limited subsets of galaxies are summarized in Table \ref{tab:numbers}. 

\begin{table}
\begin{center}
\begin{tabular}{|l||c|c|c|} \hline
Fig. & Redshift & Conditions & Galaxies \\ \hline \hline
Fig. \ref{fig:ccor00to24} & $0.001 \leq z < 0.006$ & $M_r \leq -17$ & 237 \\
& $0.006 \leq z < 0.012$ & & 730  \\
& $0.012 \leq z < 0.018$ & & 2221  \\
& $0.018 \leq z < 0.024$ & & 7331  \\ \hline
Fig. \ref{fig:color1} & $0.006 \leq z < 0.012$ & Red & 339  \\
& & Blue & 391  \\
& $0.012 \leq z < 0.018$ & Red & 1071  \\
& & Blue & 1150  \\ \hline
Fig. \ref{fig:mag1chance} & $0.006 \leq z < 0.012$ & $-17 < M_r \leq -15$ & 1387  \\
& & $-19 < M_r \leq -17$ & 585  \\
& & $M_r \leq -19$ & 145  \\ 
& $0.012 \leq z < 0.018$ & $-17 < M_r \leq -16$ & 1425  \\
& & $-19 < M_r \leq -17$ & 1575  \\
& & $M_r \leq -19$ & 646  \\ \hline
Fig. \ref{fig:type1chance} & $0.006 \leq z < 0.012$ & Early & 103 \\
& & Late & 627 \\
& $0.012 \leq z < 0.018$ & Early & 458 \\
& & Late & 1763 \\ \hline
\end{tabular}
\caption{Summary of SDSS galaxy subsamples used for analyses in this study. Parameter sets without the description of the range of $M_r$ adopt $M_r \leq -17$. The colors of galaxies are defined as $M_g - M_r > 0.6$ ({\it red}) and $M_g - M_r \leq 0.6$ ({\it blue}). The morphology of galaxies is determined by $c_{\rm inv} \leq 0.35$ ({\it early}) and $c_{\rm inv} > 0.35$ ({\it late}). See text for details.}
\label{tab:numbers}
\end{center}
\end{table}

\begin{figure}
\begin{center}
\includegraphics[width=0.55\linewidth]{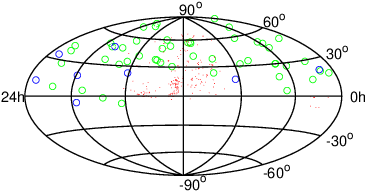} \hfill 
\includegraphics[width=0.41\linewidth]{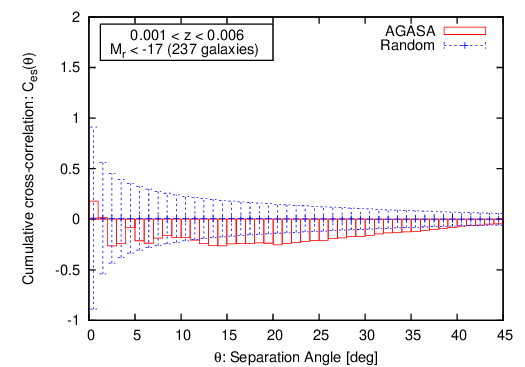} \\
\includegraphics[width=0.55\linewidth]{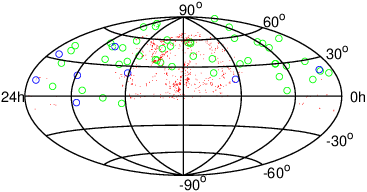} \hfill 
\includegraphics[width=0.41\linewidth]{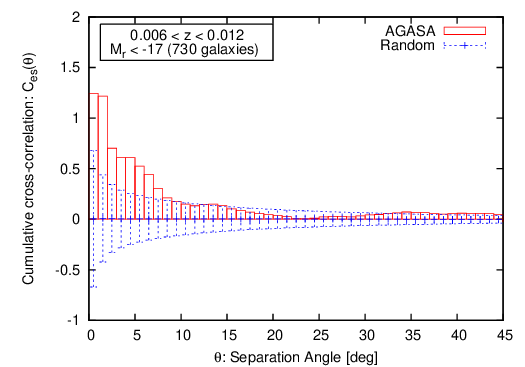} \\
\includegraphics[width=0.55\linewidth]{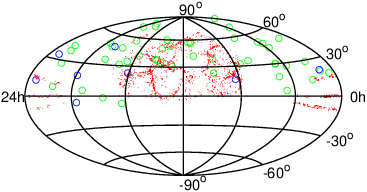} \hfill 
\includegraphics[width=0.41\linewidth]{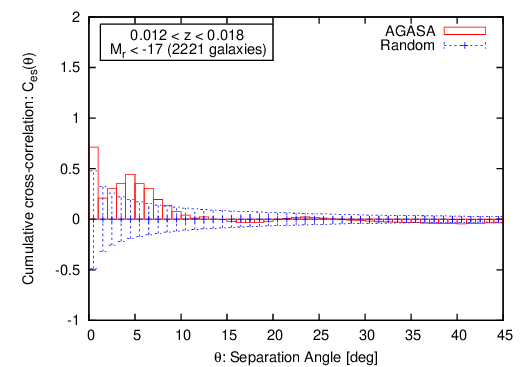} \\
\includegraphics[width=0.55\linewidth]{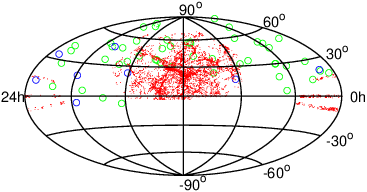} \hfill 
\includegraphics[width=0.41\linewidth]{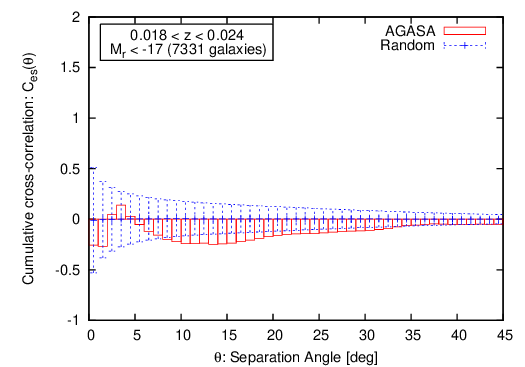} 
\caption{{\it left}: Skymaps of the AGASA events ($E > 4 \times 10^{19}$ eV; {\it green circles}, $E > 10^{20}$ eV; {\it blue circles}) and SDSS galaxies with $M_r \leq -17$ ({\it red points}). The redshift ranges of SDSS galaxies are $0.001 \leq z < 0.006$, $0.006 \leq z < 0.012$, $0.012 \leq z < 0.018$, and $0.018 \leq z < 0.024$ from top to bottom. {\it right}: Cumulative cross-correlation functions between the AGASA events and SDSS galaxies shown in the corresponding left panels ({\it histogram}). We also show the averages ({\it points}) and standard deviations ({\it bars}) of the cumulative cross-correlation functions calculated from 100000 realizations of randomly distributed events in blue.} 
\label{fig:ccor00to24}
\end{center}
\end{figure}

\subsection{Statistical analysis} \label{stat}

A lot of tests of anisotropy in the arrival distribution of HECRs focus on the excess of the number of events within a circle centered by source candidates over random events. We adopt the same approach, although the statistical quantities used in this study is different. We analyzed the PAO and AGASA data by using a differential cross-correlation function which took the anisotropic apertures of these experiments and galaxy surveys into account.\cite{takami09c} \ The effect of their boundary should be also taken with care in this study, since areas observed by the SDSS and the AGASA are limited and non-uniform. Here, we adopt a {\it cumulative} cross-correlation function defined similarly to the differential cross-correlation function as 
\begin{equation}
C_{\rm es}(\theta) = \frac{EG(<\theta) - EG'(<\theta) - E'G(<\theta) + E'G'(<\theta)}{E'G'(<\theta)}. 
\label{eq:ccor}
\end{equation}
Here, $E$, $G$, $E'$ and $G'$ represent HECR events, galaxies in a sample, HECR events randomly put with number density proportional to the detector aperture, and galaxies randomly put following the angular selection function (e.g., survey window, bright star mask, and so on) in the observed sky, respectively. $EG(<\theta)$ is the normalized number of pairs between $E$ and $G$ within the angular distance of $\theta$. For normalization, the number of the pairs are divided by the product of the number of HECR events and the number of galaxies. $EG'(<\theta)$, $E'G(<\theta)$, and $E'G'(<\theta)$ are defined similarly to $EG(<\theta)$. $E'$ and $G'$ enable us to correct the non-uniformities of the HECR aperture and the galaxy sampling. This function allows us to investigate the excess of HECR events around galaxies over random events. We can also define the (angular) cumulative auto-correlation functions of galaxies similarly to Eq. \ref{eq:ccor} as 
\begin{equation}
C_{\rm g}(\theta) = \frac{GG(<\theta) - 2GG'(<\theta) + G'G'(<\theta)}{G'G'(<\theta)}. 
\label{eq:acor}
\end{equation}
These functions were originally introduced by Ref.~\citen{landy93} for a differential auto-correlation function and by Ref.~\citen{blake06} for a differential cross-correlation function.

The aperture of a ground array depends on the declination of observed directions reflecting the daily rotation of the Earth. The declination dependence of the exposure (= aperture $\times$ observation time) can be analytically estimated as,\cite{sommers01}
\begin{equation}
\omega(\delta) \propto \cos(a_0) \cos (\delta) \sin (\alpha_m) 
+ \alpha_m \sin (a_0) \sin (\delta), 
\label{eq:exp}
\end{equation}
where $\alpha_m$ is given by 
\begin{equation}
\alpha_m = \left\{
\begin{array}{ll}
0 & {\rm if}~\xi > 1 \\
\pi & {\rm if}~\xi < -1 \\
\cos^{-1}(\xi) & {\rm otherwise}
\end{array}
\right., 
\end{equation}
and 
\begin{equation}
\xi \equiv \frac{\cos (\Theta) - \sin (a_0) \sin (\delta)}{\cos (a_0) \cos (\delta)}, 
\end{equation}
if observation time is sufficiently larger than a day. Here, $a_0$ is the terrestrial latitude of a ground array and $\Theta$ is the zenith angle for a data cut because of experimental reasons. For the AGASA, $a_0 = 35^{\circ}47'$ and $\Theta = 45^{\circ}$.\cite{hayashida00} \ The $E'$ sample in Eq.~\ref{eq:ccor} is generated following Eq.~\ref{eq:exp}. We set the number of $E'$ events to be 200,000 in order that the distribution of random events reflects the AGASA aperture sufficiently.

The survey area of the SDSS is also limited and complicated due to masks. The SDSS observes about one fifth of the whole sky and the depth of observations slightly depends on directions. We take account of these effects by distributing random galaxies (i.e., $G'$ in Eq.~\ref{eq:ccor}) following the angular selection function. We use {\tt mangle} software\cite{hamilton04,swanson08} to do so. We distribute at least twenty times more random galaxies than the real ones, which is an usual choice. We confirmed that these particular choices of the number of random events and random galaxies do not affect the results.

The significance of positive correlation is estimated by comparing a cumulative cross-correlation function calculated from the AGASA events to that calculated from randomly distributed events with the same number of events as that of the AGASA events. We generate 100000 realizations of random events and calculate a cumulative cross-correlation function for every realization. Then, we count the number of realizations in which the value of the cumulative cross-correlation function is larger than that calculated from the AGASA data at every $\theta$, and the number is divided by 100000. This number gives the probability that the correlation realized by the AGASA data within $\theta$ is realized from randomly distributed events by chance. However, this probability is not a real chance probability when we scan the probability over $\theta$. Since the scan is regarded as trials, the chance probability includes a trial factor. Thus, in order to estimate the true detection probability of the positive correlation, we must compensate this probability by a trial factor. In other words, the probability must be penalized.\cite{tinyakov01,tinyakov04,finley04} \ In Section \ref{redshift}, \ref{color}, and \ref{parameters}, we search for the potential signals of the positive correlation and discuss their dependence on the properties of galaxies with non-penalized probability. Then, we estimate penalty factors in all the cases tested in the three sections, and evaluate the true detection probabilities of the positive correlation in Section \ref{penalty}. Then, we will check the dependence discussed in the former three sections based on the penalized probabilities. Note that since the other parameters are set {\it a priori}, we need not consider statistical penalty from them.

\section{Results} \label{results}

\subsection{Dependence on redshift} \label{redshift}

The left panels of Fig. \ref{fig:ccor00to24} show the positions of galaxies ({\it dots}) and the arrival directions of the AGASA events ({\it circles}) for the four redshift ranges of galaxies: $0.001 \leq z < 0.006$, $0.006 \leq z < 0.012$, $0.012 \leq z < 0.018$, and $0.018 \leq z < 0.024$ from top to bottom, for visibility. Events with energies beyond $10^{20}$ eV, are shown as blue circles. The distributions of galaxies are highly structured; we can find the filamentary structures and clusters of galaxies. We see that more than ten events correlate with the filamentary structures by eye, despite that only about a half of events remains inside the region where the SDSS covers. On the other hand, the AGASA events do not correlate with Virgo cluster and Coma cluster, which are seen at around the center of the top panel and the bottom panel, respectively. The lack of the correlation with Virgo cluster was already pointed out both for the PAO events\cite{gorbunov08} and for the AGASA events.\cite{takami09c}

Whether the AGASA events globally correlate with galaxy distribution or not can be quantified by cumulative cross-correlation functions shown in the right panels, which correspond to the left panels. For comparison, we also show the cumulative cross-correlation functions between events randomly distributed with being compatible to the geometry of the AGASA exposure and the same galaxies with error bars. We simulated the same number of random events as the AGASA data 100000 times and calculated a cross-correlation function for each event set. Then, we estimated the averages ({\it crosses}) and standard deviations of the cross-correlation functions ({\it error bars}). These standard deviations can be regarded as 1 $\sigma$ statistical errors if the fluctuation of a cumulative cross-correlation function follows the Gaussian distribution. The cumulative cross-correlation functions calculated from the data are in excess of the cumulative cross-correlation functions predicted by the randomly distributed event sets at small angular scale in the cases of $0.006 \leq z < 0.012$ and $0.012 \leq z < 0.018$. On the other hand, positive excess is not found for $0.001 \leq z < 0.006$ and $0.018 \leq z < 0.024$ at small angular scale. 

\begin{wrapfigure}{l}{6.6cm}
\centerline{\includegraphics[width=0.98\linewidth]{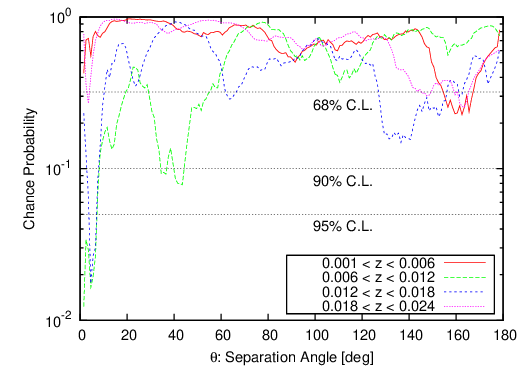}}
\caption{Probabilities that the positive correlations of the AGASA events and SDSS galaxy sets within the 4 redshift ranges is reproduced from randomly distributed events by chance as a function of $\theta$. Dotted lines represent 68\% C.L., 90\% C.L., and 95\% C.L..}
\label{fig:zchance}
\end{wrapfigure}

The significance of these positive correlations in an angular bin can be estimated by using the method described in Section \ref{stat}. Fig. \ref{fig:zchance} shows the chance probabilities that the correlation signals of the AGASA data are produced from randomly distributed events for $\theta = 2^{\circ}, 3^{\circ}, \cdots, 179^{\circ}$. Note that $\theta = 1^{\circ}$ and $180^{\circ}$ are not considered because the chance probabilities in these bins are physically meaningless. The accuracy to determine the arrival directions of HECRs above $4 \times 10^{19}$ eV by the AGASA is larger than $1^{\circ}$ ($1.8^{\circ}$\cite{takeda99}). A cumulative cross-correlation function is $0$ at $\theta = 180^{\circ}$ by definition. The chance probabilities are less than 5\% at small angular scale ($\leq 7^{\circ}$) in $0.006 \leq z < 0.012$ and $0.012 \leq z < 0.018$. These are the potential signals of positive correlation which reflect the fact that $\sim 10$ events seem to correlate with the filamentary structure of galaxies in the left panels of Fig. \ref{fig:ccor00to24}. In the following two sections, we focus on galaxies in these two redshift ranges and test the dependence of the cross-correlation on the properties of galaxies, i.e., color, $r$-band absolute magnitude, and morphology. 

\subsection{Dependence on the color of galaxies} \label{color}

\begin{figure}
\begin{center}
\includegraphics[width=0.48\linewidth]{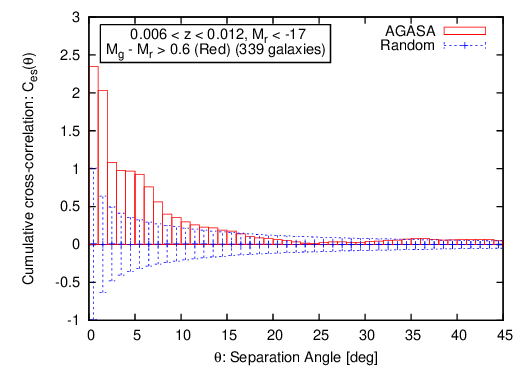} \hfill 
\includegraphics[width=0.48\linewidth]{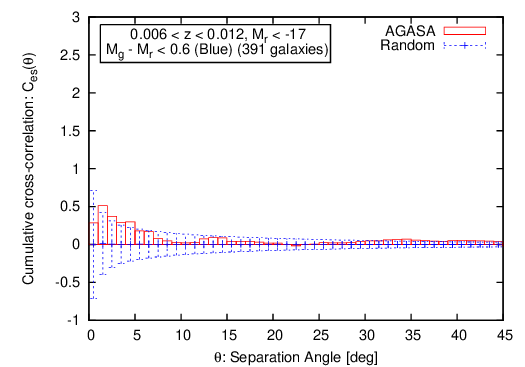} 
\caption{Cumulative cross-correlation functions between the AGASA events and red ({\it left}) and blue ({\it right}) galaxies with $0.006 \leq z < 0.012$ ({\it histogram}). The points and error bars are the average and standard deviation of cumulative cross-correlation functions calculated from random events.}
\label{fig:color1}
\end{center}
\end{figure}

\begin{figure}
\begin{center}
\includegraphics[width=0.48\linewidth]{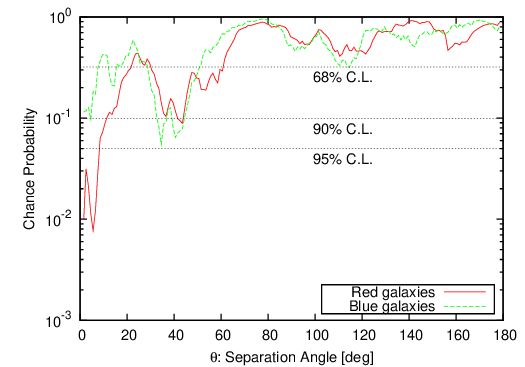} \hfill
\includegraphics[width=0.48\linewidth]{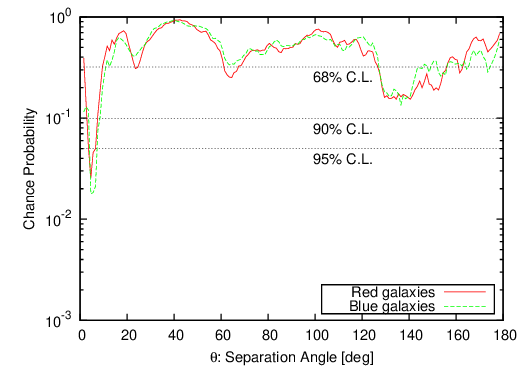} 
\caption{Same as Fig. \ref{fig:zchance}, but for red ($M_g - M_r > 0.6$; red) and blue ($M_g - M_r \leq 0.6$; green) galaxies with $0.006 \leq z < 0.012$ ({\it left}) and $0.012 \leq z < 0.018$ ({\it right}).}
\label{fig:col1chance}
\end{center}
\end{figure}

Fig. \ref{fig:color1} shows the cumulative cross-correlation functions of the AGASA events with red ({\it left}) and blue ({\it right}) galaxies within $0.006 \leq z < 0.012$. The cumulative cross-correlation functions calculated from random events are also plotted. The cross-correlation function of the data is in excess of that of random events in the case of red galaxies, while the data is consistent with random events for blue galaxies within error bars. This fact can be also observed in the plot of chance probabilities (the left panel of Fig. \ref{fig:col1chance}). The probability that the observed positive correlation is realized from randomly distributed events by chance is less than 5\% at small angular scale for the red galaxies. The same analysis is applied to galaxies within $0.012 \leq z < 0.018$ (the right panel of Fig. \ref{fig:col1chance}), but the difference between red and blue galaxies is small, although both the chance probabilities are less than 5\% at $\sim 5^{\circ}$.

\subsection{Dependence on other galaxy parameters} \label{parameters}

Following the former section, we test the dependence on the other two galaxy parameters, $r$-band absolute magnitude and morphology.

\begin{figure}
\begin{center}
\includegraphics[width=0.48\linewidth]{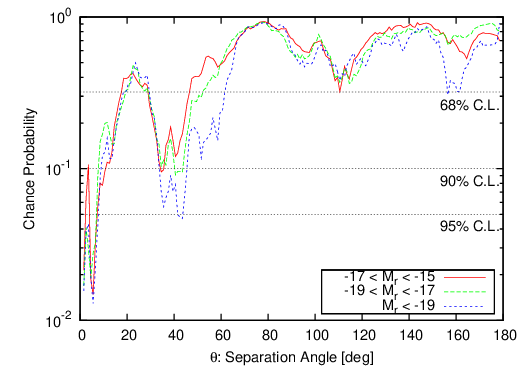} \hfill
\includegraphics[width=0.48\linewidth]{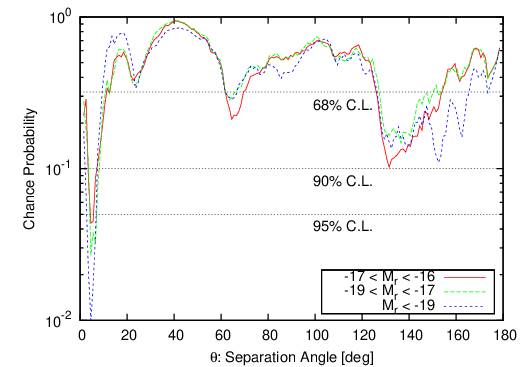}
\caption{{\it left}: Same as Fig. \ref{fig:zchance}, but for galaxies with $-17 < M_r \leq -15$ ({\it red}), $-19 < M_r \leq -17$ ({\it green}), and $M_r \leq -19$ ({\it blue}) in the redshift range of $0.006 \leq z < 0.012$. {\it right}: Same as Fig. \ref{fig:zchance}, but for galaxies with $-17 < M_r \leq -16$ ({\it red}), $-19 < M_r \leq -17$ ({\it green}), and $M_r \leq -19$ ({\it blue}) in the redshift range of $0.012 \leq z < 0.018$.} 
\label{fig:mag1chance}
\end{center}
\end{figure}

Fig. \ref{fig:mag1chance} is the same as Fig. \ref{fig:col1chance}, but for the dependence on $r$-band absolute magnitude of galaxies within $0.006 \leq z < 0.012$ ({\it left}) and $0.012 \leq z < 0.018$ ({\it right}), respectively. In the left panel, all the three galaxy samples show the chance probability less than 5\% at small angular scale, but the difference among them is small. On the other hand, the chance probabilities seem to be smaller for brighter galaxies in the right panel. 

\begin{figure}
\begin{center}
\includegraphics[width=0.48\linewidth]{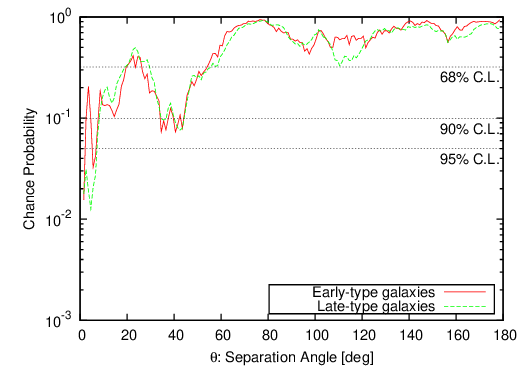} \hfill
\includegraphics[width=0.48\linewidth]{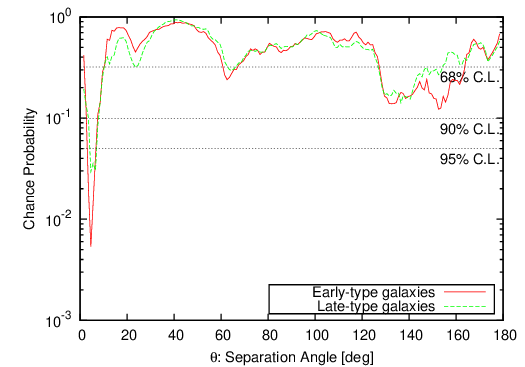} 
\caption{Same as Fig. \ref{fig:zchance}, but for early-type ({\it red}) and late-type ({\it blue}) galaxies with $0.006 \leq z < 0.012$ ({\it left}) and $0.012 \leq z < 0.018$ ({\it right}).}
\label{fig:type1chance}
\end{center}
\end{figure}

Next, we consider the morphology dependence of the cross-correlation. Fig. \ref{fig:type1chance} is the same as Fig. \ref{fig:mag1chance}, but the original galaxy samples are divided into early-type galaxies and late-type galaxies. In the range of $0.006 \leq z < 0.012$ ({\it left}), the chance probability for late-type galaxies is smaller than that for early-type galaxies at small angular scale, but not so clear. On the other hand, a positive correlation signal of early-type galaxies is stronger than that of late-type galaxies in the range of $0.012 \leq z < 0.018$ ({\it right}). 

\subsection{Penalized Probability} \label{penalty}

We investigated the potential signals of the positive correlation in the former three sections by scanning the chance probability that the correlation signals of the AGASA events are realized from randomly distributed events by chance over angular distance. These angular scanning means that we made 178 statistical trials by different angular scale. Thus, it is possible that more than 95\% C.L. correlation signals discussed in the former three sections result from the fluctuation of randomly distributed events. In this case, the significance of the correlation signals is expected to be lower in reality. In this section, we estimate the true significance of the potential positive correlation signals found in the former three sections taking trial factors (penalty factors) into account following the methods discussed in literature (e.g., Refs.~\citen{tinyakov01,finley04}). 

\begin{wrapfigure}{l}{6.6cm}
\centerline{\includegraphics[width=0.95\linewidth]{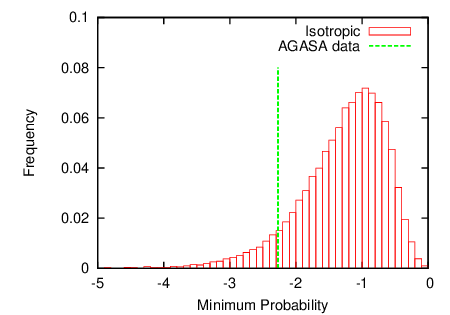}} 
\caption{Distribution of $p_{\rm min}$ calculated from randomly distributed events ({\it red}). As for galaxies to test correlation, early type galaxies located at $0.012 \leq z < 0.018$ (used in Fig. \ref{fig:type1chance}) are adopted. The green line is $p_{\rm min}$ obtained by the AGASA events. In this case, the probability that the observed correlation appears from randomly distributed events by chance is $9.5 \times 10^{-2}$ (see also Table \ref{tab:ppenalized}). }
\label{fig:pmin}
\end{wrapfigure}

The probability that a signal $p_{\rm min}$, which is the minimum chance probability in each analysis performed in the former three sections, is produced from isotropic distribution by chance (statistical fluctuations) can be estimated as follows. We generate a set of isotropically distributed HECR events with the same number as the number of the AGASA events. To this event set, we perform the same analysis as in the former three sections and derive the minimum chance probability $p_{{\rm min},1}$. In general, the angular scale to give $p_{{\rm min},1}$ is different from that of $p_{\rm min}$. If $p_{{\rm min},1} < p_{\rm min}$, this is the case where randomly distributed events could produce the observed signals by fluctuation. We repeat this process 50000 times and count the number of realizations in which $p_{{\rm min},i} < p_{\rm min}$ ($i=1,2,\cdots,50000$), then which is divided by 50000. This is the probability that the observed positive correlation is realized by fluctuations. In other words, one minus this probability is the true significance of the detection of positive correlation. We confirmed that the 50000 trials were enough to estimate the penalized significance accurately. A histogram of the distribution of $p_{{\rm min},i}$ is shown in Fig. \ref{fig:pmin} to help our understanding. A galaxy sample used in this figure is the same as that used in the right panel of Fig. \ref{fig:type1chance}. The green line, located at the left side of the peak, is the minimum chance probability $p_{\rm min}$ estimated in the right panel of Fig. \ref{fig:type1chance}. The area in the left side of the green line corresponds to the probability that $p_{\rm min}$ is realized by chance.

The penalized probabilities of all the cases investigated in the former three sections are tabulated on Table \ref{tab:ppenalized}. We can find that the strongest correlation appears for early-type galaxies within $0.012 \leq z < 0.018$ (90\% C.L.). The other penalized chance probabilities are more than 10\%, but the difference of the penalized probabilities between the properties of galaxies indicates the dependence of the correlation on the properties. The dependence found in the table is similar to the results from the discussions using non-penalized probabilities in the former three sections. For dependence on redshifts (discussed in Fig. \ref{fig:ccor00to24}), penalized significance is clearly higher in $0.006 \leq z < 0.012$ and $0.012 \leq z < 0.018$ than in the others. For the dependence on the color of galaxies, red galaxies correlates with the AGASA events stronger than blue galaxies in the redshift range of $0.006 \leq z < 0.012$. Brighter galaxies produce larger significance of the correlation in $0.012 \leq z < 0.018$, whereas the dependence is not clear at $0.006 \leq z < 0.012$. The correlation with early-type galaxies is stronger in $0.012 \leq z < 0.018$, but the morphology dependence is not clear in $0.006 \leq z < 0.012$. Thus, the dependence of the correlation signals on the properties of galaxies is not common in the redshift ranges, and therefore unclear at present. However, as discussed in the next section, each population of HECR source candidates has characteristic host galaxies. The dependence searches of the positive correlation will be a powerful tool to find HECR sources as the number of detected HECR events increases. 

\begin{table}
\begin{center}
\begin{tabular}{|l||c|c|c|} \hline
Fig. & Redshift & Conditions & Penalized $p$ \\ \hline \hline
Fig. \ref{fig:ccor00to24} & $0.001 \leq z < 0.006$ & $M_r \leq -17$ & 0.94 \\
& $0.006 \leq z < 0.012$ & & 0.23  \\
& $0.012 \leq z < 0.018$ & & 0.23  \\
& $0.018 \leq z < 0.024$ & & 0.94  \\ \hline
Fig. \ref{fig:color1} & $0.006 \leq z < 0.012$ & Red & 0.17  \\
& & Blue & 0.65  \\
& $0.012 \leq z < 0.018$ & Red & 0.31  \\
& & Blue & 0.19  \\ \hline
Fig. \ref{fig:mag1chance} & $0.006 \leq z < 0.012$ & $-17 < M_r \leq -15$ & 0.27  \\
& & $-19 < M_r \leq -17$ & 0.29  \\
& & $M_r \leq -19$ & 0.28  \\ 
& $0.012 \leq z < 0.018$ & $-17 < M_r \leq -16$ & 0.45  \\
& & $-19 < M_r \leq -17$ & 0.32  \\
& & $M_r \leq -19$ & 0.15  \\ \hline
Fig. \ref{fig:type1chance} & $0.006 \leq z < 0.012$ & Early & 0.30 \\
& & Late & 0.24 \\
& $0.012 \leq z < 0.018$ & Early & $9.5 \times 10^{-2}$ \\
& & Late & 0.34 \\ \hline
\end{tabular}
\caption{Penalized probabilities. The meaning of the left three column is the same as in Table \ref{tab:numbers}.}
\label{tab:ppenalized}
\end{center}
\end{table}

Finally, we comment on another possible statistical penalty originating from the number of trial catalogue discussed in literature.\cite{tinyakov04} \ If we test the correlation between HECRs and the matter distribution of the Universe, the statistical penalty also should be considered because a galaxy catalog is one realization of the matter distribution. In other words, for instance, it is regarded as two (dependent) trials to use two galaxy catalogue. On the other hand, this paper focuses on the correlation between HECRs and galaxies, and examines the dependence of the correlation on the various intrinsic properties of galaxies by using galaxy subsamples constructed from a priori criteria, instead of searching for the possible strongest signal of the correlation by repeating trials over different galaxy subsamples. Thus, an additional statistical penalty is not necessary. Accordingly, our results claim not the correlation with the matter distribution but the correlation with the galaxies. 

\section{Discussion} \label{discussion}

In Section \ref{redshift}, we found that the cumulative cross-correlation functions between AGASA events and the SDSS samples of galaxies in $0.006 \leq z < 0.012$ and $0.012 \leq z < 0.018$ are in excess of those calculated from random events within $\sim 10^{\circ}$. The strongest potential signals also appeared at $\sim 5^{\circ}$, although the penalized significance of these signals is not so large (77\%). These imply small deflections ($\sim 5^{\circ}$) during the propagation of HECRs.

HECRs are inevitably affected by a GMF. A GMF consists of a coherent component following the spiral arm of Milky Way with turbulent components,\cite{beck01} though the details of its structure is controversial at present.\cite{han08} \ For protons, the coherent component dominantly contributes to the deflections of HECRs due to its long coherent length, and can weaken the correlation between HECRs and source distribution because the spiral structure is generally independent of the distribution of galaxies in extragalactic space. A bisymmetric GMF model widely-adopted for HECR propagation in Galactic space\cite{stanev97} predicts deflections less than $10^{\circ}$ for protons above $10^{19.6}$ eV\cite{alvarez02,takami08b} and that a typical angular scale at which HECR-to-source correlation is expected to be $\sim 5^{\circ}$,\cite{takami09f} \ which is comparable with the angular scale giving the potential correlation signals. Thus, the positive cross-correlations support proton-dominated composition, which is consistent with the result of the HiRes, a HECR observatory in the northern hemisphere.

\begin{wrapfigure}{l}{6.6cm}
\centerline{\includegraphics[width=0.95\linewidth]{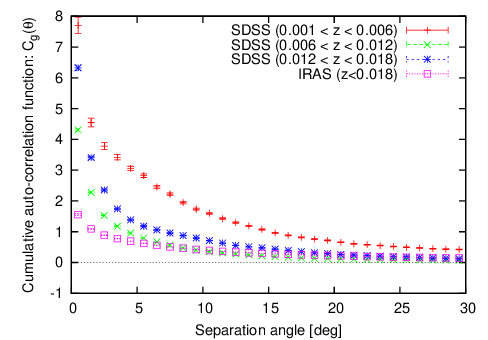}} 
\caption{Angular cumulative auto-correlation functions of galaxies in our samples used in Fig. \ref{fig:ccor00to24}. For comparison, the cumulative auto-correlation function of IRAS galaxies adopted in Ref.~\citen{takami09c} is also shown. The error bars are estimated from random distribution following the observed area of the SDSS and IRAS.}
\label{fig:acor}
\end{wrapfigure}

The signals of the cross-correlation functions obtained in Section \ref{redshift} is more positive than the result of our previous analysis using IRAS galaxies, in which the cross-correlation function of the AGASA events is consistent with that calculated from random events within $1\sigma$ error.\cite{takami09c} \ Although depending on several difference between the galaxy samples, e.g., wavebands, flux limits, observed directions, the stronger correlation signal possibly originates from galaxies at small angular scale. Fig. \ref{fig:acor} compares the cumulative angular auto-correlation functions of galaxies used in Fig. \ref{fig:ccor00to24} and the same function of a galaxy sample used in our previous study. The error bars are estimated from random event generation by putting random galaxies with the same number as the number of the real galaxies following the observed area of the SDSS and IRAS 100 times and then by estimating the standard deviation of 100 auto-correlation functions as $1\sigma$ errors. Note that we adopt these criteria on redshifts to directly compare this study with the previous study although the redshift ranges are different. This figure shows that the SDSS samples are concentrated at small angular scale stronger than the IRAS sample. These positive auto-correlations at small angular scale could be positively working.

The deflection of HECRs by IGMFs depends on their propagation distance. For a random IGMF, the total deflection angles of HECRs, $\varphi (\epsilon, D)$, are proportional to the square root of their propagation distance (e.g., Ref.~\citen{waxman95b}), 
\begin{eqnarray}
\varphi (\epsilon, D) \simeq 2.5^{\circ} Z {\epsilon_{20}}^{-1} 
{D_2}^{1/2} {B_{-9}}^{1/2} {l_{c,1}}, 
\end{eqnarray}
where $\epsilon_{20} = \epsilon / 10^{20}$ eV, $D_2 = D / 100$ Mpc, $B_9 = B / 10^{-9}$ G and $l_{c,1} = l_c / 1$ Mpc are the energy of HECRs, their propagation distance, the strength of the IGMF, the coherent length of the IGMF, respectively. The trajectories of HECRs from a more distant source are deflected stronger. The cumulative cross-correlation of the AGASA events and galaxies with $0.018 \leq z < 0.024$ consistent with that of random events could be interpreted as the effect of the IGMF as well as the lack of correlation with IRAS galaxies with $z > 0.018$ shown in Ref.~\citen{takami09c}. On the other hand, this interpretation does not apply to $0.001 \leq z < 0.006$, the nearest galaxy sample. This implies that there is not strong HECR sources within $z = 0.006$ ($\sim 25$ Mpc in the concordance cosmology) in the field observed by the SDSS.

IGMFs could trick us in some cases and could generate spurious correlation even for light-nuclei dominated composition.\cite{kotera08} \ In this scenario, HECRs can be scattered toward the Earth by strongly magnetized structures, e.g., clusters of galaxies, lobes of radio galaxies, and therefore the arrival directions of HECRs do not necessarily point out the positions of their sources although the correlation between HECRs and galaxy distribution can be detected. Ref.~\citen{kotera08} reported that about 50\% of the PAO events published in Ref.~\citen{abraham08} could be such spurious correlations. A similar phenomena is reported by Ref.~\citen{ryu09} based on simulations of propagation of protons taking a sophisticated and simulated IGMF model\cite{das08} into account. The latter also show that predicted angles between the arrival directions of protons and AGN-like objects defined in their simulations are consistent with the correlation angles derived by the PAO\cite{abraham08} even if there are true sources far from the AGN-like objects. If the spurious correlation is dominantly realized in the Universe, it is difficult to identify HECR sources by the arrival directions of HECRs. Note that such a phenomenon strongly depends on IGMF modelling including large uncertainty at present. IGMF models by Refs.~\citen{dolag05} and \citen{takami06} predict smaller deflections of HECRs and show spurious correlation much less than the former two IGMF models. The latter IGMF model even predicts the correlation between HECRs and their sources in local Universe within a few degree for proton-dominated composition.\cite{takami08a}

The nature of galaxies which correlates with HECR events is a hint of HECR sources if the spurious correlation is not realized. Based on this motivation, we investigated the dependence of the correlation between the AGASA events and galaxies on galaxy parameters, such as color, $r$-band absolute magnitude, and morphology. We adopted $M_g - M_r$ to divide galaxies into red galaxies and blue galaxies. This value is a good indicator of contents of galaxies because the 4000 \AA~break of galaxies is included in $g$-band. This spectral break is a characteristic feature of long-lived stars like K-type stars. Thus, strong star forming and a number of young (massive) stars are not expected in galaxies with a strong spectral break at $\sim 4000$ \AA, i.e. relatively red galaxies for this color definition. The dependence of the correlation of HECRs and galaxies on their colors informs us of the epochs of HECR generation in the history of galaxies and environment which includes HECR sources. We summarize the current understanding of the host galaxies of HECR source candidates, such as GRBs, magnetars, and AGN.

GRBs (at least long-duration ones) are thought to be related to explosions of massive stars at the end of their life.\cite{galama98,hjorth03,stanek03} \ Many observations have indicated that the typical nature of GRB host galaxies is faint star-forming galaxies dominated by a young stellar population\cite{christensen04} and tends low luminosities,\cite{fruchter99} \ low masses,\cite{lefloch03} \ and therefore low metallicities.\cite{gorosabel05,wiersema07,levesque10} \ Reflecting these facts, the colors of GRB host galaxies tend to be blue \cite{lefloch03}, even in nearby Universe.\cite{sollerman05} \ Note that the definition of the color of galaxies in Ref.~\citen{lefloch03}, $R - K$, is essentially the same as the $M_g - M_r$ color, because both are indicators of old (long-lived) star population in galaxies.  Morphologically, GRBs correlate with irregular or late-type galaxies,\cite{lefloch03} \ but a recent study which use the largest sample of GRB host galaxies reports no compelling evidence that GRB host galaxies are peculiar galaxies in the point of their morphologies.\cite{savaglio09}

We should take care of the time-delay of HECRs during their propagation in intergalactic space when we discuss whether GRBs are HECR sources or not from the nature of galaxies which correlate with HECR events. The color of GRB host galaxies is expected to become red after massive stars are exploded. In other words, the host galaxies are no longer blue at present if the time-delay is larger than the lifetime of massive stars. The masses of GRB progenitors are thought to be larger than $\sim 25 M_{\odot}$ or typically $\sim 40 M_{\odot}$ to require black hole formation by stellar core collapse under a collapser model.\cite{woosley93} \ The typical main sequence lifetime of such massive stars is $\sim 10^7$ yr. On the other hand, the time-delay of HECRs depends on the modelling of IGMF. 
In a uniformly turbulent IGMF, which is a simple case, 
the time-delay, $\tau_{\rm d}(\epsilon, D)$ is estimated as 
\begin{eqnarray}
\tau_{\rm d}(\epsilon, D) &\sim& 10^5 Z 
{\epsilon_{20}}^{-2} {D_2}^2 {B_{-9}}^2 l_{c,1} ~~~{\rm yr},
\end{eqnarray}
and this allows us to observe blue host galaxies. However, IGMF is structured in the Universe. Regions in which have stronger magnetic fields can produce larger time-delay. Several IGMF models predict the time-delay of HECRs sufficiently smaller than the lifetime of massive stars.\cite{dolag05,takami06} \ If IGMF indicated by these works is realized in the Universe, the discussion on colors of the correlated galaxies does not lose its efficiency. Other IGMF models predict that a significant fraction of HECRs has time-delay larger than $\sim 10^7$ yr even for protons.\cite{sigl04,das08} \ In this case, since the trajectories of such HECRs are highly deflected, the host galaxies of GRBs becomes already red and also correlation between HECRs and their sources itself can disappear. Thus, it is difficult to confirm the GRB origin of HECRs from correlation studies of HECRs if these strong IGMF models are realized.

The production mechanism of magnetars is still controversial, but magnetars are mainly thought to be produced through core-collapse of massive stars because a large fraction of magnetar candidates correlate with supernova remnants or massive stellar cluster\cite{gaensler01} (see Ref.~\citen{mereghetti08} for a review). All the magnetar candidates are detected inside or very close to our galaxy, and therefore the nature of the host galaxies of magnetars is also poorly known. Based on the hypothesis that magnetars are generated by core-collapse of massive stars, they are often produced in star-forming galaxies. GRB060218 (sometimes categorized into X-ray flush) is a possible candidate of a GRB leaving a magnetar, classified into low-luminosity GRBs, and the progenitor mass is estimated as $\sim 20 M_{\odot}$,\cite{mazzali06} \ which is smaller than that of GRBs which leave black holes (high-luminosity GRBs). It was theoretically predicted that low-luminosity GRBs accompany the production of magnetars.\cite{toma07} \ Thus, the host galaxies of newly born magnetars could be redder than that of high-luminosity GRBs. Note that low-luminosity GRBs themselves are also possible to accelerate particles up to the highest energies.\cite{murase06}

AGN which are theoretically motivated as HECR source candidates are classified into radio galaxies, which are dominated by non-thermal radiation. The host galaxies of such AGN have been well investigated and have early-type morphology and red color.\cite{krolik00}

The strongest signal of the positive correlation obtained in this study originated from early-type galaxies at $0.012 \leq z < 0.018$. This signal indicates that some kinds of AGN activity are related to HECR generation among known source candidates. However, we could not find clear common dependence between the two redshift ranges, e.g., the AGASA events correlated with red galaxies stronger tha blue galaxies in $0.006 \leq z < 0.012$ while this was not clear at $0.012 \leq z < 0.018$. Thus, we cannot make a clear conclusion of HECR sources at present. The discussion above was based on the results from very limited detected events. Increasing the number of detected events can allow us to obtain a clear conclusion. 

\begin{figure}
\begin{center}
\includegraphics[width=0.48\linewidth]{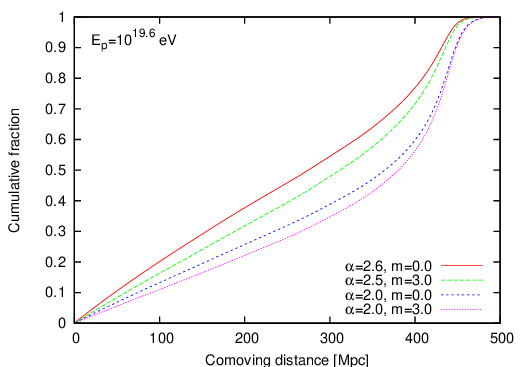} 
\includegraphics[width=0.48\linewidth]{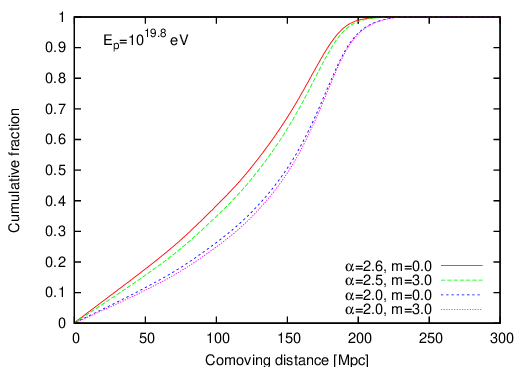} 
\caption{Cumulative fractions of the flux of protons arriving from sources with distances represented as the horizontal axis to the total flux of protons observed at the Earth with energies of $10^{19.6}$ eV ({\it left}) and $10^{19.8}$ eV ({\it right}), respectively. HECR sources are assumed to be distributed uniformly. $\alpha$ is a spectral index of HECR emission and $m$ is a index of a power-law cosmological evolution scenario proportional to $( 1 + z )^m$. }
\label{fig:cumcont}
\end{center}
\end{figure}

Finally, we discuss a fraction for HECRs emitted from sources at a distance to contribute to the total flux. Fig. \ref{fig:cumcont} shows the cumulative fraction of the HE proton flux arriving from sources with distances represented as the horizontal axis to the total flux observed at the Earth. We assume that HECR sources are distributed uniformly and a $\Lambda$CDM cosmology with $H_0 = 71$ km s$^{-1}$ Mpc$^{-1}$, $\Omega_m = 0.3$, and $\Omega_{\Lambda} = 0.7$, as the Hubble parameter, the ratio of matter density to the critical density, and the ratio of cosmological constant to the critical density, respectively. These cumulative fractions are calculated under the continuous energy-loss approximation of Bethe-Heitler pair creation\cite{chodorowski92} and photopion production.\cite{mucke00} \ The cumulative fractions are calculated under several assumptions on the nature of HECR sources; a spectral index $\alpha$ of HECR emission spectrum and a cosmological evolution factor of HECR source density with a shape of $(1 + z)^m$, and rectilinear propagation (neglecting IGMFs). We can see that the fraction of protons arriving at the Earth from sources within 100 Mpc is only 10-20 \% for protons with energies of $10^{19.6}$ eV ({\it left}). Even for protons with $10^{19.8}$ eV ({\it right}), the fraction is maximally $\sim$ 40 \%. However, IGMFs could more deflect the trajectories of HECRs emitted from more distant sources and therefore the cross-correlation signal between arriving HECRs and distant sources may be weaken as discussed above. Therefore, it is natural to focus on the cross-correlation of HECRs and nearby galaxy distribution. In addition, sources in redshift for every 0.006, which corresponds to every $\sim 25$ Mpc, can equally contribute to the total HECR flux up to $z = 0.024$ because the cumulative functions can be approximated by straight lines below $\sim 100$ Mpc. 

\section{Conclusion} \label{conclusion}

In this study, we investigated the correlation between the arrival directions of HECRs detected in the northern sky (the AGASA data) and volume-limited samples of galaxies observed by the SDSS. We found that the excess of the cumulative cross-correlation functions between the AGASA events and SDSS galaxies over random HECR events at small angular scale in the redshift ranges of $0.006 \leq z < 0.012$ and $0.012 \leq z < 0.018$. These cross-correlation signals imply light composition at the highest energy. Then, we examined the dependence of the correlation on the properties of galaxies in these two redshift ranges. The properties of galaxies correlating with HECRs can give us a hint of HECR sources unless the spurious correlation is dominated, as galaxies hosting HECR sources are expected to be characteristic features (discussed in Section \ref{discussion}). The strongest correlation signal originated from early-type galaxies in $0.012 \leq z < 0.018$ at 90\% C.L. (penalized significance) and the AGASA events have excess over random events at small angular scale for red galaxies rather than blue galaxies at $0.006 \leq z < 0.012$, but we could not find common dependence between the two redshift ranges probably due to the limitation of the number of HECR events. In order to obtain a clear conclusion, the more number of HECR events should be adopted for the analysis. Data accumulation in the northern sky is in progress. The total exposure of the Telescope Array have already reached 75\% of the AGASA exposure\cite{taketa09} and is expected to reach that of the AGASA in this year. In the future, the extreme statistics by projected HECR observatories, e.g., Extreme Universe Space Observatory (JEM-EUSO)\cite{ebisuzaki09} and the northern site of the PAO\cite{harton09} will help more to overcome the statistics problem and the method discussed in this paper will give us a significant constraint on HECR sources.

\section*{Acknowledgments}

We are grateful to I.~Kayo and K.~Maeda for useful discussions. This work is supported by Grants-in-Aid for Scientific Research from the Ministry of Education, Culture, Sports, Science and Technology (MEXT) of Japan through No.19104006 (K.S.).The work of T.N. is supported by the JSPS fellow. This work is also supported by World Premier International Research Center Initiative (WPI Initiative), from the MEXT of Japan.

Funding for the SDSS and SDSS-II has been provided by the Alfred P. Sloan Foundation, the Participating Institutions, the National Science Foundation, the U.S. Department of Energy, the National Aeronautics and Space Administration, the Japanese Monbukagakusho, the Max Planck Society, and the Higher Education Funding Council for England. The SDSS Web Site is http://www.sdss.org/.

The SDSS is managed by the Astrophysical Research Consortium for the Participating Institutions. The Participating Institutions are the American Museum of Natural History, Astrophysical Institute Potsdam, University of Basel, University of Cambridge, Case Western Reserve University, University of Chicago, Drexel University, Fermilab, the Institute for Advanced Study, the Japan Participation Group, Johns Hopkins University, the Joint Institute for Nuclear Astrophysics, the Kavli Institute for Particle Astrophysics and Cosmology, the Korean Scientist Group, the Chinese Academy of Sciences (LAMOST), Los Alamos National Laboratory, the Max-Planck-Institute for Astronomy (MPIA), the Max-Planck-Institute for Astrophysics (MPA), New Mexico State University, Ohio State University, University of Pittsburgh, University of Portsmouth, Princeton University, the United States Naval Observatory, and the University of Washington.


\begin{thebibliography}{99}
\bibitem{waxman95} E.~Waxman, \PRL{75,1995,386}.
\bibitem{vietri95} M.~Vietri, \AJ{453,1995,883}.
\bibitem{murase06} K.~Murase, K.~Ioka, S.~Nagataki and T.~Nakamura, \JL{Astrophys.~J.~Lett.,651,2006,L5}.
\bibitem{takahara90} F.~Takahara, \PTP{83,1990,1071}.
\bibitem{rachen93} J.~Rachen and P.~Biermann, \JL{Astron. \& Astrophys.,272,1993,161}.
\bibitem{berezhko08} E.~Berezhko, \JL{Astrophys.~J.~Lett.,684,2008,L69}.
\bibitem{farrar09} G.~Farrar and A.~Gruzinov, \AJ{693,2009,329}.
\bibitem{takami11} H.~Takami and S.~Horiuchi, \JL{Astropart.~Phys.,34,2011,749}
\bibitem{pe'er09} A.~Pe'er, K.~Murase, and P.~M\'{e}sz\'{a}ros, \PRD{80,2009,123018}.
\bibitem{arons03} J.~Arons, \AJ{589,2003,871}.
\bibitem{murase09} K.~Murase, P.~M\'{e}sz\'{a}ros and B.~Zhang, \PRD{79,2009,103001}.
\bibitem{inoue06} S.~Inoue, G.~Sigl, M.~Francesco, Proc.~30th~Int.~Cosmic~Ray~Conf., \textbf{4} (2008), 555.
\bibitem{kang97} H.~Kang, J.~Rachen, P.~Biermann, \JL{Mon.~Not.~Roy.~Astron.~Soc.,286,1997,257}.
\bibitem{greisen66} K.~Greisen, \PRL{16,1966,748}.
\bibitem{zatsepin66} G.~Zatsepin and V.~Kuz'min, \JL{J.~Exp.~Theor.~Phys.~Lett.,4,1966,78}.
\bibitem{takeda99} M.~Takeda et~al., \AJ{522,1999,225}.
\bibitem{veron06} M.~-P.~Veron-Cetty and P.~Veron, \JL{Astron. \& Astrophys,455,2006,773}.
\bibitem{abraham07} J.~Abraham et~al., \JL{Science,318,2007,938}.
\bibitem{abraham08} J.~Abraham et~al., \JL{Astropart.~Phys.,29,2008,188}.
\bibitem{george08} M.~R.~George, A.~C.~Fabian, W.~H.~Baumgartner, R.~F.~Mushotzky and J.~Tueller, \JL{Mon.~Not.~Roy.~Astron.~Soc.,388,2008,L59}.
\bibitem{ghisellini08} G.~Ghisellini, G.~Ghirlanda, F.~Tavecchio, F.~Fraternali, G.~Pareschi, \JL{Mon.~Not.~Roy.~Astron.~Soc.,390,2008,L88}.
\bibitem{kashti08} T.~Kashti and E.~Waxman, \JL{J.~Cosmo.~Astropart.~Phys.,05,2008,006}.
\bibitem{takami09c} H.~Takami, T.~Nishimichi, K.~Yahata and K.~Sato, \JL{J.~Cosmo.~Astropart.~Phys.,06,2009,031}.
\bibitem{aublin09} J.~Aublin et~al., Proc.~31th~Int.~Cosmic Ray Conf. (2009), Pub. ID 491.
\bibitem{hague09} J.~D.~Hague et~al., Proc.~31th~Int.~Cosmic~Ray~Conf. (2009), Pub. ID 143.
\bibitem{takami08b} H.~Takami and K.~Sato, \AJ{681,2008,1279}.
\bibitem{takami09f} H.~Takami and K.~Sato, \AJ{724,2009,1456}.
\bibitem{abraham10} J.~Abraham et~al, \PRL{104,2010,091101}.
\bibitem{abbasi05} R.~Abbasi et~al., \AJ{622,2005,910}.
\bibitem{abbasi10b} R.~Abbasi et~al., \PRL{104,2010,161101}.
\bibitem{stanev95} T.~Stanev, P.~L.~Biermann, J.~Lloyd-Evans, J.~P.~Rachen and A.~A.~Watson, \PRL{75,1995,3056}.
\bibitem{uchihori00} Y.~Uchihori, M.~Nagano, M.~Takeda, M.~Teshima, J.~Lloyd-Evans and A.~A.~Watson, \JL{Astropart.~Phys.,13,2000,151}.
\bibitem{saunders00} W.~Saunders et~al., \JL{Mon.~Not.~Roy.~Astron.~Soc.,317,2000,55}.
\bibitem{abbasi08b} R.~Abbasi et~al., \JL{Astropart.~Phys.,30,2008,175}.
\bibitem{abbasi10} R.~Abbasi et~al., \JL{Astrophys.~J.~Lett.,713,2010,L64}.
\bibitem{york00} D.~York et~al., \JL{Astron.~J.,120,2000,1579}.
\bibitem{smialkowski02} A.~Smialkowski, M.~Gillar M. and W.~Michalak, \JL{J.~Phys.~G.,28,2002,1359}.
\bibitem{singh04} S.~Singh, C.~-P.~Ma and J.~Arons, \PRD{69,2004,063003}.
\bibitem{gorbunov02} D.~S.~Gorbunov, P.~G.~Tinyakov, I.~I.~Tkachev and S.~V.~Troitsky, \JL{Astrophys.~J.~Lett.,577,2002,L93}.
\bibitem{tinyakov01} P.~G.~Tinyakov and I.~I.~Tkachev, \JL{J.~Exp.~Theor.~Phys.,74,2001,445}.
\bibitem{tinyakov02} P.~G.~Tinyakov and I.~I.~Tkachev, \JL{Astropart.~Phys.,18,2002,165}.
\bibitem{hague07} J.~D.~Hague, J.~A.~J.~Matthew, B.~R.~Becker and M.~S.~Gold, \JL{Astropart.~Phys.,27,2007,134}.
\bibitem{gorbunov05} D.~S.~Gorbunov and S.~V.~Troitsky, \JL{Astropart.~Phys.,23,2005,175}.
\bibitem{koers09} H.~B.~J.~Koers and P.~Tinyakov, \JL{Mon.~Not.~Roy.~Astron.~Soc.,399,2009,1005}.
\bibitem{hayashida00} N.~Hayashida et~al., \JL{Astron.~J.,120,2000,2190}.
\bibitem{takeda98} M.~Takeda et~al., \PRL{81,1998,1163}.
\bibitem{bhattacharjee00} P.~Bhattacharjee and G.~Sigl, \PRP{327,2000,109}.
\bibitem{abbasi08} R.~Abbasi et~al., \PRL{100,2008,101101}.
\bibitem{abraham08b} J.~Abraham et~al., \PRL{101,2008,061101}.
\bibitem{abazajian09} K.~N.~Abazajian et~al., \JL{Astrophys.~J.~Supp.,182,2009,543}.
\bibitem{blanton05} M.~Blanton et~al., \JL{Astron.~J.,129,2005,2562}.
\bibitem{shimasaku01} K.~Shimasaku et~al., \JL{Astron.~J.,122,2001,1238}.
\bibitem{landy93} S.~D.~Landy and A.~S.~Szalay, \AJ{412,1993,64}.
\bibitem{blake06} C.~Blake, A.~Pope, D.~Scott and B.~Mobasher, \JL{Mon.~Not.~Roy.~Astron.~Soc.,368,2006,732}.
\bibitem{sommers01} P.~Sommers, \JL{Astropart.~Phys.,14,2001,271}.
\bibitem{hamilton04} A.~J.~S.~Hamilton and M.~Tegmark, \JL{Mon.~Not.~Roy.~Astron.~Soc.,349,2004,115}.
\bibitem{swanson08} M.~E.~C.~Swanson, M.~Tegmark, A.~J.~S.~Hamilton and J.~C.~Hill, \JL{Mon.~Not.~Roy.~Astron.~Soc.,387,2008,1391}.
\bibitem{finley04} C.~B.~Finley and S.~Westerhoff, \JL{Astropart.~Phys.,21,2004,359}.
\bibitem{tinyakov04} P.~G.~Tinyakov and I.~I.~Tkachev, \PRD{69,2004,128301}.
\bibitem{gorbunov08} D.~Gorbunov, P.~Tinyakov, I.~Tkachev and S.~Troitsky, \JL{J. Exp. Theor. Phys. Lett.,87,2008,461}.
\bibitem{beck01} R.~Beck, \JL{Space~Sci.~Rev.,99,2001,243}.
\bibitem{han08} H.~Men, K.~Ferriere and J.~L.~Han, \JL{Astron. \& Astrophys.,486,2008,819}.
\bibitem{stanev97} T.~Stanev, \AJ{479,1997,290}.
\bibitem{alvarez02} J.~Alvarez-Mu\~{n}iz J., R.~Engel and T.~Stanev, \AJ{572,2002,185}.
\bibitem{waxman95b} E.~Waxman and J.~Miralda-Escude, \AJ{472,1995,L89}.
\bibitem{kotera08} K.~Kotera and M.~Lemoine, 2008, \PRD{77,2008,123003}.
\bibitem{ryu09} D.~Ryu, H.~Kang and S.~Das, Proc.~31th~Int.~Cosmic~Ray~Conf. (2009), Pub. ID 839.
\bibitem{das08} S.~Das, H.~Kang, D.~Ryu and J.~Cho, \AJ{682,2008,29}.
\bibitem{dolag05} K.~Dolag, D.~Grasso, V.~Springel and I.~Tkachev, \JL{J. Cosmo. and Astrophys.,01,2005,009}.
\bibitem{takami06} H.~Takami, H.~Yoshiguchi and K.~Sato, \AJ{639,2006,803} (Erratum, \AJ{653,2006,1583}.
\bibitem{takami08a} H.~Takami and K.~Sato, \AJ{678,2008,606}.
\bibitem{galama98} T.~J.~Galama et~al., \JL{Nature,395,1998,670}.
\bibitem{hjorth03} J.~Hjorth et~al., \JL{Nature,423,2003,847}.
\bibitem{stanek03} K.~Z.~Stanek et~al., \JL{Astrophys.~J.~Lett.,591,2003,L17}.
\bibitem{christensen04} L.~Christensen, J.~Hjorth and J.~Gorosabel, \JL{Astron. \& Astrophys.,425,2004,913}.
\bibitem{fruchter99} A.~S.~Fruchter et~al., \JL{Astrophys.~J.~Lett.,519,1999,L13}.
\bibitem{lefloch03} E.~Le Floc'h et~al., \JL{Astron. \& Astrophys.,400,2003,499}.
\bibitem{gorosabel05} J.~Gorosabel et~al., \JL{Astron. \& Astrophys.,444,2005,711}.
\bibitem{levesque10} E.~M.~Levesque, E.~Berger, L.~J.~Kewley and M.~M.~Bagley, \JL{Astron.~J.,139,2010,694}.
\bibitem{wiersema07} K.~Wiersema et~al., \JL{Astron. \& Astrophys.,464,2007,529}.
\bibitem{sollerman05} J.~Sollerman, G.~\"{O}stlin, J.~P.~U.~Fynbo, J.~Hjorth, A.~Fruchter and K.~Pedersen, \JL{New~Astron.,11,2005,103}.
\bibitem{savaglio09} S.~Savaglio, K.~Glazebrook and D.~Le Borgne, \AJ{691,2009,182}.
\bibitem{woosley93} S.~E.~Woosley, \AJ{405,1993,273}.
\bibitem{sigl04} G.~Sigl, F.~Miniati, T.~A.~Ensslin, \PRD{70,2004,043007}.
\bibitem{gaensler01} B.~M.~Gaensler, P.~O.~Slane, E.~V.~Gotthelf and G.~Vasisht, \AJ{559,2001,963}.
\bibitem{mereghetti08} S.~Mereghetti, \JL{Astron.~\& Astrophys.~Rev.,15,2008,225}.
\bibitem{mazzali06} P.~A.~Mazzali et~al., \JL{Nature,442,2006,1018}.
\bibitem{toma07} K.~Toma, K.~Ioka, T.~Sakamoto and T.~Nakamura, \AJ{659,2007,1420}.
\bibitem{krolik00} J.~H.~Krolik , \textit{Active Galactic Nuclei}, (Princeton University Press, New Jersey, 2001).
\bibitem{chodorowski92} M.~J.~Chodorowski, A.~A.~Zdziarski A.A. and M.~Sikora, \AJ{400,1992,181}.
\bibitem{mucke00} A.~M\"{u}cke, R.~Engel, J.~P.~Rachen, R.~J.~Protheroe and T.~Stanev, \JL{Comp.~Phys.Comm.,124,2000,290}.
\bibitem{taketa09} A.~Taketa et~al., Proc.~31th~Int.~Cosmic~Ray~Conf. (2009), Pub. ID 855.
\bibitem{ebisuzaki09} T.~Ebisuzaki et~al., Proc.~31th~Int.~Cosmic~Ray~Conf. (2009), Pub. ID 1035.
\bibitem{harton09} J.~Harton et~al., Proc.~31th~Int.~Cosmic~Ray~Conf. (2009), Pub. ID 458.
\end{thebibliography}
\end{document}